\documentclass[twocolumn,prd,unsortedaddress,superscriptaddress,showpacs,a4paper,nofootinbib]{revtex4}
\usepackage{epsfig}
\def\beq{\begin{equation}}
\def\enq{\end{equation}}
\def\beqa{\begin{eqnarray}}
\def\enqa{\end{eqnarray}}
\def\MeV{\nobreak\,\mbox{MeV}}
\def\GeV{\nobreak\,\mbox{GeV}}

\def\qq{\lag\bar{q}q\rag}

\def\ss{\lag\bar{s}s\rag}
\def\mix{\lag\bar{q}g\si.Gq\rag}
\def\mixs{\lag\bar{s}g\si.Gs\rag}
\def\Gd{\lag g^2G^2\rag}
\def\G3{\lag g^3G^3\rag}

\def\ka{\kappa}

\def\La{\Lambda}

\def\rh{\rho}
\def\si{\sigma}

\def\al{\alpha}

\def\lb{\label}
\def\nn{\nonumber}

\def\X{X(3872)}
\def\Dso{D_{s0} (2317)}
\def\Dsi{D_{s1} (2460)}
\def\sig{\sigma(600)}
\def\kap{\kappa(800)}
\def\ao{a_0 (980)}
\def\fo{f_0 (980)}
\def\kal{{\cal K}}
\def\as{(\alpha , s)}
\newcommand{\rag}{\rangle}
\newcommand{\lag}{\langle}

\begin{document}

\title{\sc Do the QCD sum rules support four-quark states?}
\author{R.D.~Matheus}
\email{matheus@if.usp.br}
\affiliation{Instituto de F\'{\i}sica, Universidade de S\~{a}o Paulo, 
C.P. 66318, 05389-970 S\~{a}o Paulo, SP, Brazil}
\author{F.S.~Navarra}
\email{navarra@if.usp.br}
\affiliation{Instituto de F\'{\i}sica, Universidade de S\~{a}o Paulo, 
C.P. 66318, 05389-970 S\~{a}o Paulo, SP, Brazil}
\author{M.~Nielsen}
\email{mnielsen@if.usp.br}
\affiliation{Instituto de F\'{\i}sica, Universidade de S\~{a}o Paulo, 
C.P. 66318, 05389-970 S\~{a}o Paulo, SP, Brazil}
\author{R.~Rodrigues da Silva}
\email{romulo@df.ufcg.edu.br}
\affiliation{Universidade Federal de Campina Grande, 
58.109-900 Campina Grande, PB, Brazil}

\begin{abstract}
We test the validity of the QCD  sum rules applied to the light scalar mesons,
the charmed mesons $\Dso$ and $\Dsi$,
and the $\X$  axial meson, considered as tetraquark 
states. We find that, with the studied currents, it is possible to find an 
acceptable Borel window only for the $\X$ meson. In such a  Borel 
window we have simultaneouly  a good OPE convergence and 
a pole contribution which is bigger than the continuum contribution.  We 
interpret these results as a strong argument against the assignment of a
tetraquark structure for the light scalars and the $\Dso$ and $\Dsi$ mesons.
\end{abstract}

\maketitle

%
%
\section{Introduction}
%
%

From the light scalar mesons to the heavy ``chamonium-like'' $\X$, there
are now many states that do not fit comfortably in the spectrum of 
constituent quark model predictions. The light
scalar states below $1.5 \GeV$ are too numerous to be accommodated in a 
single $q \bar{q}$ multiplet and the 
nature of these states has been a  source of controversy for over 30 years 
\cite{amsler}. The lightest nonet is
composed, in principle, by the isoscalars $\sig$ and $\fo$, the 
isodoublet $\kap$ and the isovector $\ao$. In
a naive $q \bar{q}$ assignment it is hard to explain the $f_0$ -- 
$a_0$ mass degeneracy and why $\si$ and $\ka$ are broader than the other 
two. The strange-charmed
mesons $\Dso$ and $\Dsi$ ($J^{p} = 0^{+} \mbox{and~} 1^{+}$ respectively)
are too light to fit in the quark model prediction, with the $\Dso$ lying 
about $160 \MeV$ below most predictions \cite{Swanson}.
The $\X$, with quantum numbers $J^{PC} = 1^{++}$, does not fit in the 
charmonium spectrum and presents a strong isospin violating decay, 
disfavoring a $c \bar{c}$ assignment \cite{Swanson}. 

The structure of all these states has been extensively discussed and 
many alternatives have been proposed: meson
molecules, four-quark states, glueballs (in the case of scalars) and hybrids 
($qg\bar{q}$). The idea that the light
scalar mesons could be four-quark bound states has been first proposed by 
Jaffe in 1977 \cite{Jaffe}, and has later
been extrapolated to heavier sectors. Jaffe proposed
that some states may be composed of two quarks and two antiquarks 
($qq\bar{q}\bar{q}$) arranged so that the (anti)quark-(anti)quark
correlation is important, forming what is called a (anti)diquark. Recently  
the existence 
of tetraquarks received  some support from lattice calculations \cite{suga}, 
which, however, 
are not yet definitive.

The QCD sum rules (QCDSR) \cite{svz,rry,SNB} have been previously used to 
study the light scalars \cite{leves.1st,nosso.leves,koch,CHZ}, 
the strange-charmed scalars \cite{nosso.ds,OTHERA}
and the $\X$ \cite{nosso.x} as diquark-antidiquark states. 

In \cite{nosso.leves} it was assumed that the light scalars were tetraquarks 
and 
no attempt to compute their masses in QCDSR was performed. Instead a 
calculation of their decay widths, using their experimental masses was 
presented.  At the 
same time, in \cite{nosso.ds} the masses of several charmed scalars were 
calculated.
With the tetraquark hypothesis, the decay width of the $D_{sJ}(2317)$ and of 
the $X(3872)$ 
were calculated in \cite{marina_D_decay} and in \cite{naniel} respectively.

While the masses were  often very close to the experimental values, the 
widths were not
always as narrow as found in experiments. This is expected because,  unless 
some symmetry 
violation is involved, tetraquarks can decay more easily since no quark pair 
creation is 
needed and a ``fall - apart'' decay is allowed. 

From 2003 to 2006, the QCDSR calculations of masses and decay widths evolved 
rapidly and became 
much more rigorous. While the first calculations aimed only at estimating 
some order of 
magnitude and only the Borel stability was checked, the last ones were much 
more concerned 
with OPE convergence and with pole dominance, which are traditional tests, 
from  which one can determine the quality of the calculation.  

The improvement of the standards was also motivated by  
``the pentaquark experience''. In this case from the begining there was an 
experimental 
controversy about the very existence of the particle. After the first round 
of promising 
results, it was realized \cite{MATTHEUS,mananiel} that the pentaquark sum 
rules were 
problematic, because it was always very difficult to find a Borel window in 
which one would 
have at the same time good OPE convergence and pole dominance. In favor of 
the QCDSR 
practitioners it must be said that sum rules with more than three quarks 
presents new and challeging aspects. 
The number of possible interpolating currents increases significantly and 
also one has to 
worry about subtracting the two-hadron-reducible contributions \cite{kondo}, 
a problem never encountered 
before in QCDSR calculations.  Finally, to make things even more complex, 
there may be a
mixing between two and four-quark states. This  requires the combination of 
interpolating 
fields of different dimensions with the introduction of a new parameter.

Relating pentaquarks and tetraquarks may be very instructive. In both cases 
negative results were gradually found, but there was always still a lot of 
work to 
be done, such as computing higher order contributions to the OPE, instanton 
contributions, 
$\alpha_s$ corrections and new possible interpolating  currents. Therefore 
it took a long 
time until a negative opinion about pentaquarks was formed in the QCDSR 
community.
We have now gathered evidence to believe 
that QCD sum rules calculations of tetraquark properties have reached the 
same turning point 
found before in the case of pentaquarks. This is the point where, even 
though there are 
improvements to be made, we do not believe that these improvements will 
change  the conclusion 
of a series of works pointing to the non-existence of tetraquarks.

In this work we review some of the tetraquark sum rules with special 
attention to the validity 
limits of the method. In section II we work out  the sum rules of the axial 
strange-charmed 
$\Dsi$ as a prototype for this analysis, and extend the  discussion to   
other states. The study of the $\Dsi$  complements the calculations 
published in \cite{nosso.ds}. In section III we extend the analysis of 
section I to the light scalars, studying  some of the interpolating fields 
proposed for these states  and study also the charmed scalars. In section 
IV we examine the sum rules for the $X(3872)$.

%
%
\section{The charmed axial meson $\Dsi$}
%
%

The interpolating operator for $\Dsi$ (as a diquark-antidiquark state) is 
built by extension
of the operator used to describe $\Dso$ in ref.~\cite{nosso.ds}, changing 
the diquarks so we get an axial current:
\beq
j_\mu={i\epsilon_{abc}\epsilon_{dec}\over\sqrt{2}}[(u_a^TC\gamma_5c_b)
(\bar{u}_d\gamma_\mu C\bar{s}_e^T)+u\leftrightarrow d]\;,
\label{field}
\enq
where $a,~b,~c,~...$ are color indices and $C$ is the charge conjugation
matrix. We choose to work with an axial light antidiquark to avoid
instanton contributions to the sum rule \cite{hjlee}.

The sum rule for the charmed axial meson is constructed from the two-point
correlation function:
\beqa
\Pi_{\mu\nu}(q) &=& i\int d^4x ~e^{iq.x}\lag 0
|T[j_\mu(x)j^\dagger_\nu(0)]
|0\rag
= \nn \\ & = & 
-\Pi_1(q^2)(g_{\mu\nu}q^2-{q_\mu q_\nu})+\Pi_0(q^2)q_\mu
q_\nu.
\lb{2po}
\enqa
Since the axial vector current is not conserved, the two functions,
$\Pi_1$ and $\Pi_0$, appearing in Eq.~(\ref{2po}) are independent and
have respectively the quantum numbers of the spin 1 and 0 mesons.

The calculation of the
phenomenological side proceeds by inserting intermediate states for
the axial vector meson and parametrizing its coupling to the current 
$j_\mu$, in Eq.~(\ref{field}), in terms
of the meson decay constant $f_{D_{s1}}$ as:
\beq\label{eq: decay}
\lag 0 |
j_\mu|D_{s1}\rag =\sqrt{2}f_{D_{s1}} m_{D_{s1}}^4\epsilon_\mu~, 
\enq
the phenomenological side
of Eq.~(\ref{2po}) can be written as 
\beq
\Pi_{\mu\nu}^{phen}(q^2)={2f_{D_{s1}}^2m_{D_{s1}}^8\over
m_{D_{s1}}^2-q^2}\left(-g_{\mu\nu}+ {q_\mu q_\nu\over m_{D_{s1}}^2}\right)
+\cdots\;, \lb{phe} \enq
where the Lorentz structure $g_{\mu\nu}$ projects out the spin 1 state.  The 
dots denote higher axial-vector resonance contributions that will be
parametrized, as usual, through the introduction of a continuum
threshold parameter $s_0$ \cite{io1}.

In the OPE side we work at leading order and consider condensates up to 
dimension six. We deal with the strange quark as a light one and consider
the diagrams up to order $m_s$. To keep the charm quark mass finite, we
use the momentum-space expression for the charm quark propagator. We
calculate the light quark part of the correlation
function in the coordinate-space, which is then Fourier transformed to the
momentum space in $D$ dimensions. The resulting light-quark part is combined 
with the charm-quark part before it is dimensionally regularized at $D=4$.

We can write the $g_{\mu\nu}$ structure of the correlation function in the 
OPE side 
in terms of a dispersion relation:
\beq
- q^2 \Pi_1(q^2) \equiv \Pi^{OPE}(q^2)= \int_{m_c^2}^\infty ds {\rho(s)\over 
s-q^2}\;,
\lb{ope}
\enq
where the spectral density is given by the imaginary part of the correlation
function: $\rho(s)={1\over\pi}\mbox{Im}[\Pi^{OPE}(s)]$. After making a Borel
transform on both sides, and transferring the continuum contribution to
the OPE side, the sum rule for the $g_{\mu\nu}$ structure can be written as
\beq
- 2f_{D_{s1}}^2m_{D_{s1}}^8e^{-m_S^2/M^2}=\int_{m_c^2}^{s_0}ds~ 
e^{-s/M^2}~\rho(s)\;,
\lb{sr}
\enq
where $\rho(s)=\rho^{pert}(s)+\rh^{\qq}(s)+\rh^{\lag G^2\rag}
(s)+\rh^{mix}(s)+\rh^{\qq^2}(s)$, with
\beq
\rho^{pert}(s)={-1\over 2^{12} 3\pi^6}\int_\La^1 
\!\!\!\! d\al \,\,\kal^4\as 
\left({1-\al\over\al}
\right)^3(3+\al),
\lb{ope.dim0}
\enq
\beqa
\rho^{\qq}(s)={-1\over 2^{8} 3 \pi^4}\int_\La^1 
\!\!\!\! d\al \,\,\kal^2\as
~{1-\al\over\al}
\bigg[
6 m_s \bigg(4 \qq + 
\nn\\
+ (1 + \al) \ss \bigg)
+ m_c \qq \left(-{2 \over \al} + 1 + \al \right)
\bigg],
\lb{ope.dim3}
\enqa
\beqa
\rho^{\lag G^2\rag}(s)&=&{-\Gd\over 2^{12} 3^2 \pi^6}
\int_\La^1 
\!\!\!\! d\al \,\,\kal\as {1-\al\over\al}
\bigg[m_c^2 (3+\al)  \times
\nn\\&& \times \!\!
 \left({1-\al\over\al}\right)^2 \!+\! 6 \kal\as \left({1 \over \al} - 2 
\right)\bigg],
\lb{ope.dim4}
\enqa
\beqa
\rho^{mix}(s)&=&{1 \over 2^{7} 3 \pi^4} \int_\La^1 
\!\!\!\! d\al \,\,\kal\as
\bigg[
\nn\\
&& m_s \bigg(6 \mix + \mixs (2-3\al) \bigg) +
\nn\\
&-&
{2 m_c \mix \over \al^2} (1-3\al+2\al^3)
\bigg],
\lb{ope.dim5}
\enqa
\beq
\rho^{\qq^2}(s)={\qq \ss \over 12 \pi^2} \int_\La^1 
\!\!\!\! d\al \,\,\kal\as ,
\lb{ope.dim6}
\enq
where $\La=m_c^2/s$ and $\kal\as = m_c^2 - \al s$.
For the charm quark propagator with two gluons attached we used
the momentum-space expressions given in ref.~\cite{rry}.

In order to extract the mass $m_{D_{s1}}$ without knowing  about the value of
the decay constant $f_{D_{s1}}$, we take the derivative of Eq.~(\ref{sr})
with respect to $1/M^2$, divide the result by Eq.~(\ref{sr}) and
obtain:
\beq
m_{D_{s1}}^2={\int_{m_c^2}^{s_0}ds ~e^{-s/M^2}~s~\rho(s)\over
\int_{m_c^2}^{s_0}ds ~e^{-s/M^2}~\rho(s)}\;.
\lb{m2}
\enq

In the numerical analysis of the sum rules, the values used for the
quark
masses and condensates are \cite{narpdg}: 
$m_c=1.23\,\GeV$, $m_s = 0.1\,\GeV$,
$\lag\bar{q}q\rag=\,-(0.23)^3\,\GeV^3$,
$\lag\bar{q}g\si.Gq\rag=m_0^2\lag\bar{q}q\rag$ with \cite{SNB}
$m_0^2=0.8\,\GeV^2$ and 
$\lag g^2G^2\rag=0.88~\GeV^4$.
We evaluate the sum rules 
for three values of $s_0$: $\sqrt{s_0} = 2.7 \GeV$, 
$\sqrt{s_0} = 2.9 \GeV$ and $\sqrt{s_0} = 3.1 \GeV$. 

\subsection{Pole versus continuum}

\begin{figure}[h] 
\centerline{\epsfig{figure=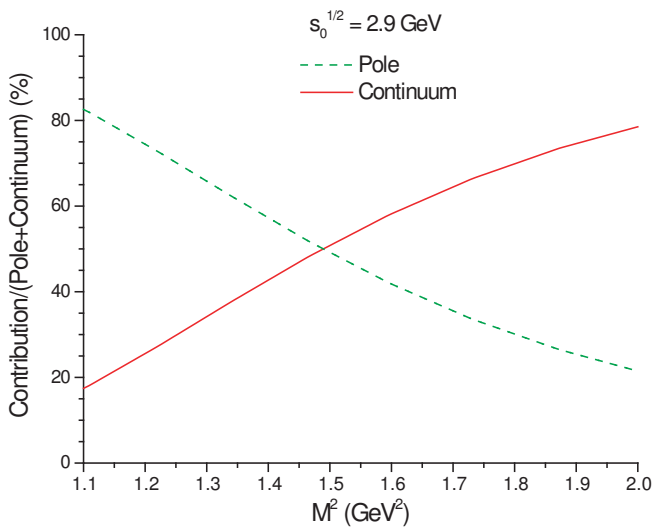,height=70mm}}
\caption{The dashed line shows the relative pole contribution (the
pole contribution divided by the total, pole plus continuum,
contribution) and the solid line shows the relative continuum
contribution.  The pole contribution should be bigger than the
continuum, which happens for $M^2 < 1.5$ GeV$^2$ for $\sqrt{s_0} = 2.9 \GeV$.}
\label{figpvc} 
\end{figure}

We get an upper limit  for $M^2$ by imposing that the QCD
continuum contribution should be smaller than the pole contribution.
The maximum value of $M^2$ for which this constraint is satisfied
depends on the value of $s_0$.  The comparison between pole and
continuum contributions for $\sqrt{s_0} = 2.9 \GeV$ is shown in
Fig.~\ref{figpvc}.

The same analysis for the other values of the continuum  
threshold gives $M^2 < 1.4$  GeV$^2$ for $\sqrt{s_0} = 2.7 \GeV$ and
$M^2 < 1.6$  GeV$^2$ for $\sqrt{s_0} = 3.1 \GeV$.

In Fig.~\ref{figmx}, we show the $D_{s1}$ mass obtained from
Eq.~(\ref{m2}), in the $M^2$ region below the upper limit 
obtained above.  We limit ourselves to the region $M^2>1.2~\GeV^2$
where the curves are more stable. Averaging the mass over all this region
we get:
\begin{figure}[h] 
\centerline{\epsfig{figure=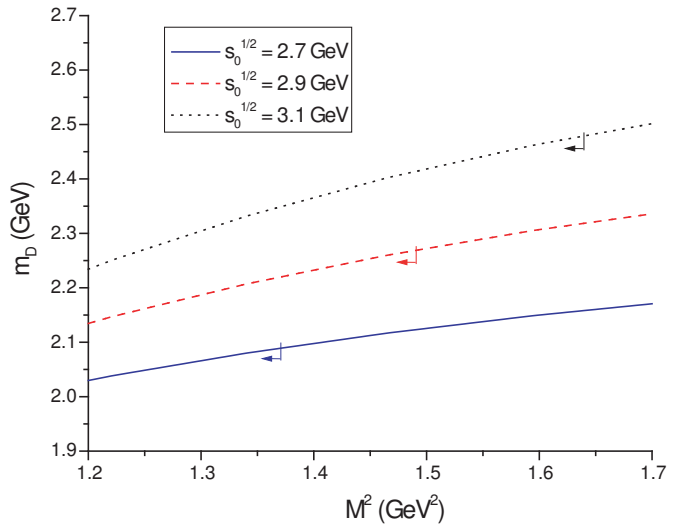,height=70mm}}
\caption{The $D_{s1}$ mass as a function of the sum rule parameter
($M^2$) for different values of the continuum threshold.  
The arrows indicate the region
allowed by the upper limit imposed by the dominance of the
QCD pole contribution.}
\label{figmx} 
\end{figure} 
\beq
m_{D_{s1}} = (2.3\pm 0.2) \GeV~,
\enq
which is compatible with the experimental value $\Dsi$ \cite{Swanson}. 

\begin{figure}[h] 
\centerline{\epsfig{figure=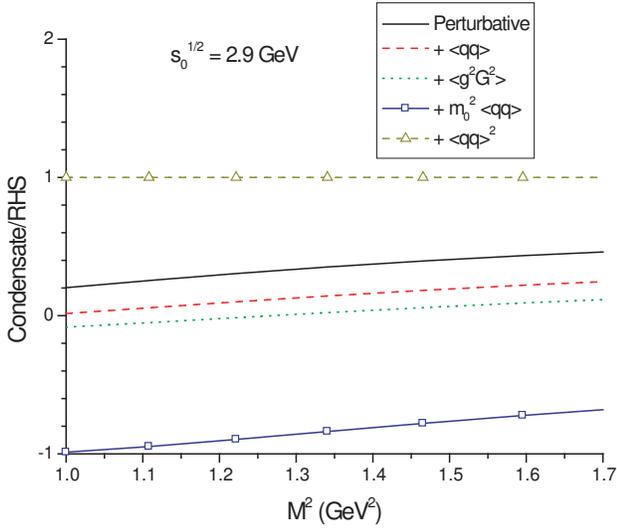,height=70mm}}
\caption{The OPE convergence in the region $1.0 \leq M^2 \leq
1.7~\GeV^2$ for $\sqrt{s_0} = 2.9 \GeV$. We start with the perturbative
contribution and each
subsequent line represents the addition of a condensate of higher 
dimension in the expansion.}
\label{figconv} 
\end{figure}

\subsection{OPE convergence}

There is however a stronger constraint to the lower bound of the $M^2$ region.
We have to analyze the convergence of the OPE by comparing the relative 
contribution of 
each term in Eqs.~(\ref{ope.dim0}) to (\ref{ope.dim6}),
to the right hand side of Eq.~(\ref{sr}). The series converges
 better for higher values of $M^2$, so that requiring a good
convergence sets a lower limit to $M^2$. This analysis in shown
in figure \ref{figconv}.

Figure \ref{figconv} shows no convergence in any region 
allowed by the upper bound given by pole/continuun analysis.
This means that the lower bound given by OPE convergence will be higher
than the upper bound, and there is no ``sum rule window'' where we can
completely trust the  results for this current.

The results above illustrate very well how we can reproduce the mass of a 
given state and then after a more carefult analysis conclude that the state 
is not a particle as such, 
being rather one of the possible continuum excitations.

%
%
\section{The scalar mesons}
%
%

\subsection{Light scalars}

The same situation described in the last section is encountered in many  
sum rules 
with interpolating operators built with more than three quark fields. The 
light 
scalar meson interpolating operators used in ref.~\cite{nosso.leves} are:
\beqa
j_\si&=&\epsilon_{abc}\epsilon_{dec}(u_a^TC\gamma_5d_b)(\bar{u}_d\gamma_5C
\bar{d}_e^T),
\nn\\
j_{f_0}&=&{\epsilon_{abc}\epsilon_{dec}\over\sqrt{2}}\left[(u_a^TC\gamma_5s_b)
(\bar{u}_d\gamma_5C\bar{s}_e^T)+u\leftrightarrow d\right],
\nn\\
j_{a_0}&=&{\epsilon_{abc}\epsilon_{dec}\over\sqrt{2}}\left[(u_a^TC\gamma_5s_b)
(\bar{u}_d\gamma_5C\bar{s}_e^T)-u\leftrightarrow d\right],
\nn\\
j_\ka&=&\epsilon_{abc}\epsilon_{dec}(u_a^TC\gamma_5d_b)(\bar{q}_d\gamma_5C
\bar{s}_e^T),\;\;\;\bar{q}=\bar{u},\bar{d}.
\label{int2}
\enqa
They yield very low upper limits to $M^2$ when the pole and
continuum contributions are analysed: $M^2 < 0.73 \, \GeV^2$ for $\ao$ and 
$\fo$, $M^2 < 0.62 \, \GeV^2$ for the $\kap$ and $M^2 < 0.54 \, \GeV^2$ for 
the $\sig$. 
\begin{figure}[h] 
\centerline{\epsfig{figure=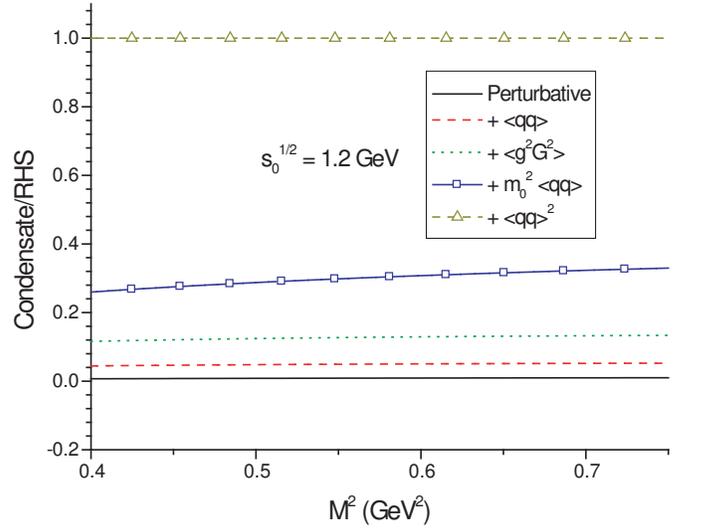,height=70mm}}
\caption{The $\ao / \fo $ OPE convergence in the region $0.4 \leq M^2 \leq
0.7~\GeV^2$ for $\sqrt{s_0} = 1.2 \GeV$.}
\label{f0a0_conv.eps} 
\end{figure}
The analysis was performed with the same parameters used in 
\cite{nosso.leves}:
$s_0^\si=1.0\,\GeV^2$, $s_0^\ka=1.2\,\GeV^2$, $s_0^{f_0}=1.5\,\GeV^2$.

In figure \ref{f0a0_conv.eps} we show the OPE convergence for 
$\ao$ and $\fo$ (which have the same sum rule), in the same way shown in
figure \ref{figconv}. In  this figure we see that there is no OPE convergence
in any region allowed by the upper bound. In fact, the situation  
of the light scalars is even worse than that  of the  $\Dsi$, since the 
relative contribution of the dimension-6 condensate is even bigger. A
possible reason for this is the fact that we are working with very small
values for the Borel Mass ($M^2<1 ~ \GeV^2$). As a matter of fact, once the 
integral on the right hand side of Eq.~\ref{ope} is evaluated, the OPE side
becomes a series with decreasing powers of $M^2$, which  eventually
become negative so that    higher condensates will be
divided by higher and higher powers of $M^2$. In the case of the
tetraquarks the series begins with $M^{10}$ and  one still has a positive
power of $M^2$ for the dimension-8 condensate. However, it is hardly 
justifiable to truncate the series at this point since  higher dimension
condensates will be proportional to $(1/M^2)^{{D -10 \over 2}}$, where $D$ 
is the dimension of the condensate and for $M^2 < 1 \GeV^2$ these condensates
will not be suppressed, at least for $D\sim10$. 

It is interesting to notice that the authors of ref.~\cite{cesa} have 
arrived at the conclusion that the $a_0(980)$ scalar meson is not a four-quark
state using a different criterion. The authors of ref.~\cite{cesa} have 
annalyzed the QCD sum rules of the $a_0(980)$ meson considered as a normal 
two-quark state, and also as a four-quark state. While they could reproduced
both the mass and width of the  $a_0(980)$ considered as a two-quark state,
they were not able to reproduce the width of the $a_0(980)$ considered as a 
four-quark state.

%
%
%

It could be argued that these problems are related with the specific currents
that we are working with, and that there could be 
other currents that might  work better. In Ref. 
\cite{CHZ}, five different interpolating operators for each
of the light scalar mesons have been tested. In the case of the $\si$ these 
currents were:
\beqa
S_3^{\si}&=&(u_a^TC\gamma_5d_b)
(\bar{u}_a\gamma_5C \bar{d}_b^T - \bar{u}_b\gamma_5C \bar{d}_a^T),
\nn\\
V_3^{\si}&=&(u_a^TC\gamma_{\mu}\gamma_5d_b)
(\bar{u}_a\gamma^{\mu}\gamma_5C \bar{d}_b^T - \bar{u}_b\gamma^{\mu}
\gamma_5C \bar{d}_a^T),
\nn\\
T_6^{\si}&=&(u_a^TC\si_{\mu\nu}d_b)
(\bar{u}_a\si^{\mu\nu}C \bar{d}_b^T - \bar{u}_b\si^{\mu\nu}C \bar{d}_a^T),
\nn\\
A_6^{\si}&=&(u_a^TC\gamma_{\mu}d_b)
(\bar{u}_a\gamma^{\mu}C \bar{d}_b^T - \bar{u}_b\gamma^{\mu}C \bar{d}_a^T),
\nn\\
P_3^{\si}&=&(u_a^TCd_b)
(\bar{u}_aC \bar{d}_b^T - \bar{u}_bC \bar{d}_a^T).
\label{curr.CHZ}
\enqa
The currents for the other light scalars can be obtained by the following
substitutions: $\ka : (ud)(\bar{u}\bar{d}) \rightarrow (ud)(\bar{d}\bar{s})$,
$f_0 : (ud)(\bar{u}\bar{d}) \rightarrow (us)(\bar{u}\bar{s}) + (ds)(\bar{d}
\bar{s})$
and $a_0 : (ud)(\bar{u}\bar{d}) \rightarrow (us)(\bar{u}\bar{s}) - (ds)
(\bar{d}\bar{s})$.
The authors of \cite{CHZ} have tested all these currents and various linear 
combinations and found out that the better results were obtained with the 
particular combination: 
$\eta^{\si}_1 = \mbox{cos}\theta A_6^{\si} + \mbox{sin}\theta V_3^{\si}$, 
with $\mbox{cos}\theta = 1 / \sqrt{2}$.
They also obtain  good results for the other light scalars with similar 
combinations.

We used the same analysis used above with the spectral densities obtained 
in \cite{CHZ} 
and agree that the OPE convergence up to dimension 8 is quite good.
On the other hand the pole dominance requirement imposes very low upper limits
to $M^2$: $M^2 < 0.8 \GeV^2$ for $a_0$ or
$f_0$ ($\sqrt{s_0} = 1.6 \GeV$), $M^2 < 0.45 \GeV^2$ for $\kappa$ 
($\sqrt{s_0} = 1.2 \GeV$)and $M^2 < 0.35 \GeV^2$ for $\sigma$
($\sqrt{s_0} = 1. \GeV$),

This means that the whole sum rule window  lies below 
$M^2 < 1 \GeV^2$ and, as commented above, it is at least dangerous
to truncate  the series at this order.

\subsection{Charmed scalars}

\begin{figure}[h] 
\centerline{\epsfig{figure=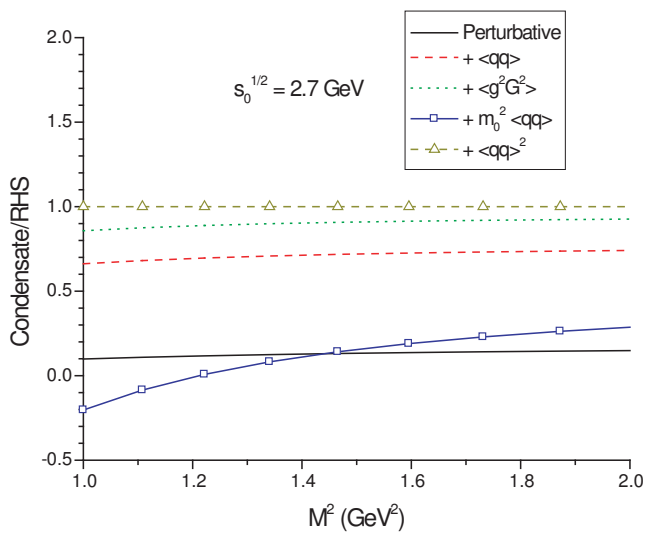,height=70mm}}
\caption{The $j_s$ OPE convergence in the region $1.0 \leq M^2 \leq
2.0~\GeV^2$ for $\sqrt{s_0} = 2.7 \GeV$.}
\label{js_conv.eps} 
\end{figure}

In the case of the charmed scalar $\Dso$,
the current used for it in ref~\cite{nosso.ds} is:
\beqa
j_s&=&{\epsilon_{abc}\epsilon_{dec}\over\sqrt{2}}\left[(u_a^TC
\gamma_5c_b)(\bar{u}_d\gamma_5C\bar{s}_e^T)+u\leftrightarrow d\right].
\label{js}
\enqa
If we require  that the pole contribution  be bigger than the continuum 
contribution we obtain  $M^2 < 1.37 \GeV^2$ for $\sqrt{s_0} = 2.7 \GeV$.
In figure \ref{js_conv.eps} we show the OPE convergence for the current  
(\ref{js}) and we see that the OPE  is still not convergent
in the allowed region, as in the case of tme meson $D_{s1}(2460)$ studied
in the previous section.

%
%
\section{Heavier tetraquarks}
%
%

\begin{figure}[h] 
\centerline{\epsfig{figure=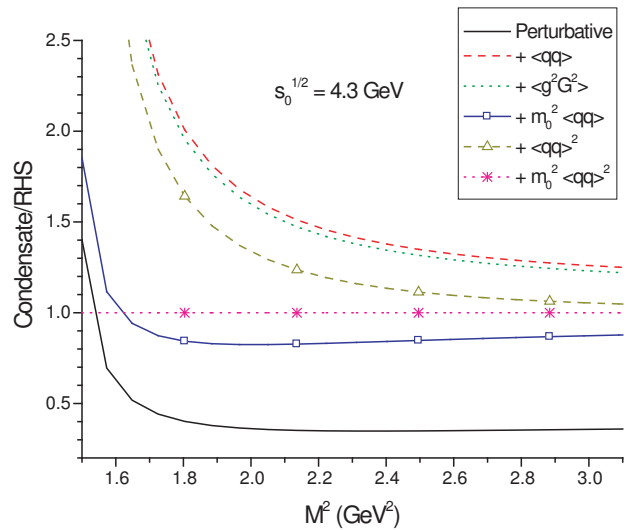,height=70mm}}
\caption{The $j^{X}_\mu$ OPE convergence in the region $1.5 \leq M^2 \leq
3.1~\GeV^2$ for $\sqrt{s_0} = 4.3 \GeV$.}
\label{x_conv.eps} 
\end{figure}

The situation improves as the quarks in the interpolating
operator become heavier. In the case of the $X(3872)$, 
the following operator was used in ref.~\cite{nosso.x}:
\beqa
j^{X}_\mu&=&{i\epsilon_{abc}\epsilon_{dec}\over\sqrt{2}}\bigg[(q_a^TC
\gamma_5c_b)(\bar{q}_d\gamma_\mu C\bar{c}_e^T) + 
\nn \\ &+& (q_a^TC\gamma_\mu c_b)
(\bar{q}_d\gamma_5C\bar{c}_e^T)].
\label{fieldX}
\enqa
The continuum contribution analysis for $j^{X}_\mu$ sets the upper limit at
$M^2 < 2.6 \, \GeV^2$ for a threshold of  $\sqrt{s_0} = 4.3 \, \GeV$. The OPE
convergence in this region is shown in figure \ref{x_conv.eps}. 
We see that, for $M^2 > 1.9 \GeV^2$, the  addition 
of a subsequent term of the expansion brings  the curve (representing the 
sum) closer to an asymptotic value (which was normalized to 1). 
Furthermore the changes in  this  curve become smaller with 
increasing dimension. These are the requirements for convergence and in this 
case we get a sum rule window in the region $1.9 \GeV^2  < M^2 < 2.6 \GeV^2$.
The mass obtained in \cite{nosso.x} considering the allowed Borel window is
\beq
m_X=(3.92\pm0.13)~\GeV\;,
\enq 
which is compatible with the experimental value $X(3872)$.

A similar situation is found if we replace the $c$ quarks in 
Eq.~(\ref{fieldX}) by $b$ quarks in order to predict the $X_b$ mass (as done 
in ref. \cite{nosso.x}). In this case the allowed Borel window is 
in the region $6.0 \GeV^2  < M^2 < 7.0 \GeV^2$, and the predicted mass is
\beq
m_{X_b}=(10.14\pm0.11)~\GeV\;,
\enq 
which is in agreement with the findings in ref.~\cite{guo}.

From what was seen above we can conclude that for heavier tetraquarks the 
sum rules satisfy the validity criteria and hence allow the determination 
of the masses of these states. However, even in the present case  we can not 
yet be very positive. Firstly because, 
as usual, the calculations might still be improved, with, for example, the 
inclusion of $\alpha_s$ corrections. Secondly because it remains very 
difficult to reproduce the $X$ narrow decay width, as shown in \cite{naniel}. 
If the $X(3872)$ is proved to be a tetraquark state, it still remains to
explain why we do not observe tetraquark states with charge different from 
$c\bar{c}$ states, such as $(cu)(\bar{c}\bar{d})$ or $(cd)(\bar{c}\bar{u})$
states, which would also have trustable QCDSR as the $X(3872)$. In this sense,
the observation of a double charmed meson ($(cc)(\bar{q}\bar{q})$), which
sum rule also obey all the convergence and pole dominance criteria \cite{tcc},
would be very important to really determine the existence of tetraquark 
states.

\section{Conclusion}

We have performed a QCD sum rules calculation of the  $\Dsi$ mass considering
 this state as a tetraquark  and reanalized  other recent similar tetraquark 
sum rules, giving  special attention to the validity criteria of the method.
We found that in the case of the lighter states, 
$\si(600)$, $\ka(800)$, $a_0(980)$, $f_0(980)$ and also in the case of the 
intermediate  $\Dso$ and $\Dsi$ states, for the currents used in 
refs.~\cite{nosso.leves,nosso.ds}, there are no values of the parameters 
$s_0$ and $M^2$ that satisfy all the desired conditions. In order to obtain 
results from the sum rules for  these states we must
abandon one or more of the conditions and choose the parameters arbitrarily.

When the interpolating operator is constructed with 
heavier quark fields the situation becomes better. We found suitable regions
for the $\X$ and its extension to the bottonic sector the $X_b$.

This problem was also present in the case of the pentaquarks \cite{MATTHEUS} 
and seems connected to the high dimension  of many-quark states interpolating 
operators, independently of the exact form of these operators. This may be 
an indication from the sum rules that light many-quark states can not be 
considered as ressonances separated from the continuum.  Heavier many-quark
states are supported by the sum rules in what concerns their masses.  However 
it is very difficult (if possible) to explain their narrow decay  widths.

While one might always argue that the so far existing calculations could  be 
improved  and the final conclusions might still change, to us at this point 
in time, this seems  unlikely.

\section*{Acknowledgements}
This work has been partly supported by FAPESP and CNPq.




\end{document}